\documentclass[reprint,prx,superscriptaddress,showpacs,nofootinbib,citeautoscrip,floatfix]{revtex4-2}

\usepackage{graphicx}
\usepackage{dcolumn}
\usepackage{bm}
\usepackage{flafter}
\usepackage{float}
\usepackage{lettrine}
\usepackage{amsmath}
\usepackage{amssymb}
\usepackage{color}
\usepackage{epsfig}
\graphicspath{{eps+pdf/}}
\makeatother

\begin{document}

\title {Tellurium sublattice instability driven amorphization in the chalcogenide   AgSbTe$_2$ under pressure}
  
\author{Baihong Sun}
\affiliation {Department of Materials Science and Engineering, Guangdong Technion-Israel Institute of Technology, Shantou 515063, China}
\affiliation{Department of Materials Science and Engineering, Technion-Israel Institute of Technology, Haifa 3200003, Israel}
\author{Zihan Zhang}
\affiliation{Department of Physics and Astronomy, Uppsala University, Uppsala,75120, Sweden}
\author{Wei Luo}
\affiliation{Condensed Matter Theory Group, Materials Theory Division, Department of Physics and Astronomy, Uppsala University, Uppsala,75120, Sweden}
\author{Sergei Grazhdannikov}
\affiliation{Department of Materials Science and Engineering, Technion-Israel Institute of Technology, Haifa 3200003, Israel}
\author{Wenting Lu}
\affiliation {Department of Materials Science and Engineering, Guangdong Technion-Israel Institute of Technology, Shantou 515063, China}
\affiliation{Department of Materials Science and Engineering, Technion-Israel Institute of Technology, Haifa 3200003, Israel}
\author{Shiyu Feng}
\affiliation {Department of Materials Science and Engineering, Guangdong Technion-Israel Institute of Technology, Shantou 515063, China}
\affiliation{Department of Materials Science and Engineering, Technion-Israel Institute of Technology, Haifa 3200003, Israel}
\author{Haikai Zou}
\affiliation {Department of Materials Science and Engineering, Guangdong Technion-Israel Institute of Technology, Shantou 515063, China}
\affiliation{Department of Materials Science and Engineering, Technion-Israel Institute of Technology, Haifa 3200003, Israel}
\author{Chenxin Wei}
\affiliation {Department of Materials Science and Engineering, Guangdong Technion-Israel Institute of Technology, Shantou 515063, China}
\affiliation{Department of Materials Science and Engineering, Technion-Israel Institute of Technology, Haifa 3200003, Israel}
\author{Martin Kunz}
\affiliation{Advanced Light Source, Lawrence Berkeley National Laboratory, Berkeley, 94720, California, USA.}
\author{Hirokazu Kadobayashi}
\affiliation{Japan Synchrotron Radiation Research Institute, Sayo, Hyogo 679-5198, Japan}
\author{Bihang Wang}
\affiliation{Deutsches Elektronen-Synchrotron, DESY, Hamburg, D-22603,Germany}
\author{Azkar Saeed Ahmad}
\affiliation{Department of Materials Science and Engineering, Guangdong Technion-Israel Institute of Technology, Shantou 515063, China}
\affiliation{Guangdong Provincial Key Laboratory of Materials and Technologies for Energy Conversion, Guangdong Technion-Israel Institute of Technology, Shantou 515063, China}
\author{Yaron Amouyal}
\affiliation{Department of Materials Science and Engineering, Technion-Israel Institute of Technology, Haifa 3200003, Israel}
\author{Rajeev Ahuja}
\email{rajeev.ahuja@physics.uu.se}
\affiliation{Condensed Matter Theory Group, Materials Theory Division, Department of Physics and Astronomy, Uppsala University, Uppsala,75120, Sweden}
\author{Elissaios Stavrou}
\email{elissaios.stavrou@gtiit.edu.cn}
\affiliation{Department of Materials Science and Engineering, Guangdong Technion-Israel Institute of Technology, Shantou 515063, China}
\affiliation{Department of Materials Science and Engineering, Technion-Israel Institute of Technology, Haifa 3200003, Israel}
\affiliation{Guangdong Provincial Key Laboratory of Materials and Technologies for Energy Conversion, Guangdong Technion-Israel Institute of Technology, Shantou 515063, China}

\begin{abstract}
Pressure provides a powerful thermodynamic route to access hidden structural states in functional materials, yet the microscopic origin of pressure-induced amorphization remains elusive in many complex chalcogenides. Here we report a detailed high-pressure structural study of AgSbTe$_2$, combining synchrotron X-ray diffraction with density functional theory and molecular dynamics calculations up to 60 GPa. We uncover a pressure-driven transformation from the ambient $R\overline{3}m$ phase to a fully disordered cubic $Im\overline{3}m$ phase, through an extended intermediate amorphous state. Enthalpy calculations reveal a near-degeneracy between the $R\overline{3}m$ and $Im\overline{3}m$ structures over a broad pressure range, dictating  amorphization. Contrary to previously speculated cation vacancies, the amorphization is governed by a pronounced displacement instability of the Te sublattice.   Remarkably, the time dependent decompression pathway controls the final structural state, resulting in either amorphous (slow decompression)  or fully crystalline (fast decompression) states, indicative of a strong counterintuitive kinetic effect. 
\end{abstract}

\maketitle

\section{Introduction}
Pressure-induced amorphization (PIA) represents a striking manifestation of structural instability in condensed matter, providing access to non-crystalline states without thermal activation. Despite extensive experimental and theoretical efforts, the microscopic mechanisms governing PIA remain incompletely understood, particularly in chemically complex systems \cite{Sharma1996,Machon2014,Arora2000}. In chalcogenide materials, PIA is frequently attributed to cation vacancies, chemical disorder, or defect-mediated collapse of the crystalline framework \cite{Sun2011,Caravati2009}. This interpretation has been reinforced by studies on phase-change and related telluride systems, where intrinsic disorder is often considered a prerequisite for amorphization \cite{Xu2019,Xu2015}. However, an alternative picture where amorphization is driven by intrinsic instabilities of specific atomic sublattices, has remained largely unexplored. Distinguishing between defect- and atomic-driven mechanisms is, therefore, essential for establishing a unified understanding of PIA in complex chalcogenides.

AgSbTe$_2$, a I-V-VI$_2$ compound, offers an ideal  system to address this question. As a chemically complex chalcogenide, it combines diverse bonding environments, and strong anharmonicity, characteristics shared with both thermoelectric and phase-change materials \cite{Schneider2010,Matsunaga2011, Taneja2024, Amouyal2013, Amouyal2014}. While AgSbTe$_2$ is widely known for its high thermoelectric performance, its crystal structure at ambient conditions has long been debated up to recently \cite{sun2025}, where the $R\overline{3}m$ space group symmetry was positively identified. Likewise, the previous high-pressure studies \cite{Sun2011,Kolobov2006}, although both reported a PIA and subsequent recrystallization, they contradicted on the overall structural evolution, mainly about the exact structure of the high-pressure phase. Kumar \textit{et al}. \cite{Kumar2005} reported a B1 ($Fm\overline{3}m$, previously suggested as the ambient conditions phase) to B2 ($Pm\overline{3}m$, partially disordered) transition with an intermediate amorphous region between  17 and 26 GPa. However, Ko \textit{et al}. \cite{Ko2014} observed a PIA at 24.6 GPa, followed by a crystallization  into an unidentified phase above 49.2 GPa. More importantly, the   observed PIA remained  poorly understood. It was speculated to be related to  Ag vacancies that caused a substantial displacement of Te atoms under pressure \cite{Kumar2005,Ko2014}, somewhat resembling the case of the defected Ge$_2$Sb$_2$Te$_5$ \cite{Sun2011,Kolobov2006}.

Here we present a comprehensive high-pressure investigation of AgSbTe$_2$ using synchrotron X-ray diffraction in combination with density functional theory (DFT) and molecular dynamics (MD) simulations, tracking its structural evolution up to 60 GPa and across multiple compression–decompression pathways. The results clearly demonstrate that AgSbTe$_2$ exhibits a  PIA at $\approx$  19.2 GPa  and remains amorphous up to  37 GPa. Upon further compression, a pressure-induced recrystallization (PIR) occurred, forming a solid solution, fully disordered, $Im\overline{3}m$ structure. The amorphous state emerges from a near-degeneracy of the enthalpies of  the $R\overline{3}m$ and the $Im\overline{3}m$ structures. We demonstrate that the observed PIA   is governed by a pronounced displacement instability of the Te sublattice, rather than by cation vacancies.  Surprisingly, it is accompanied by a strong counterintuitive  kinetic  effect upon decompression,  contrary to previously reported rate-dependent decompression mechanisms \cite{Lin2020}. These results establish a vacancy-independent pathway for amorphization in chalcogenides and provide a framework for understanding stress-driven structural disorder relevant to phase-change and thermoelectric materials.

\section{Results}

\subsection*{X-ray Diffraction}
Selected XRD patterns of AgSbTe$_2$ upon pressure increase  are shown in Fig. 1(a) and (b), for two independent experimental runs up to 50 and 60 GPa, respectively.  AgSbTe$_2$ remained in the $R\overline{3}m$ structure \cite{sun2025} up to $\approx$ 20 GPa in both runs. With further pressure increase , a clear PIA is evident, based on the appearance of broad diffraction features. Few additional Bragg peaks that are observed at the pressures that AgSbTe$_2$  becomes amorphous originate from: (a) the $Pnma$ high-pressure phase of Ag$_2$Te \cite{Zhang2015}, and (b) Ne PTM, see Fig. S2.  AgSbTe$_2$ remained in the amorphous state up to $\approx$ 37 GPa, followed by   recrystallization towards a BCC-like crystal structure. Based on our experimental results, this BCC-like crystal structure was identified as a fully disordered  $Im\overline{3}m$ (A2) phase. We postpone the discussion  about the determination of the exact crystal structure of the BCC-like phase for the discussion section. 

\begin{figure}[ht]
{\includegraphics[width=0.8\linewidth]{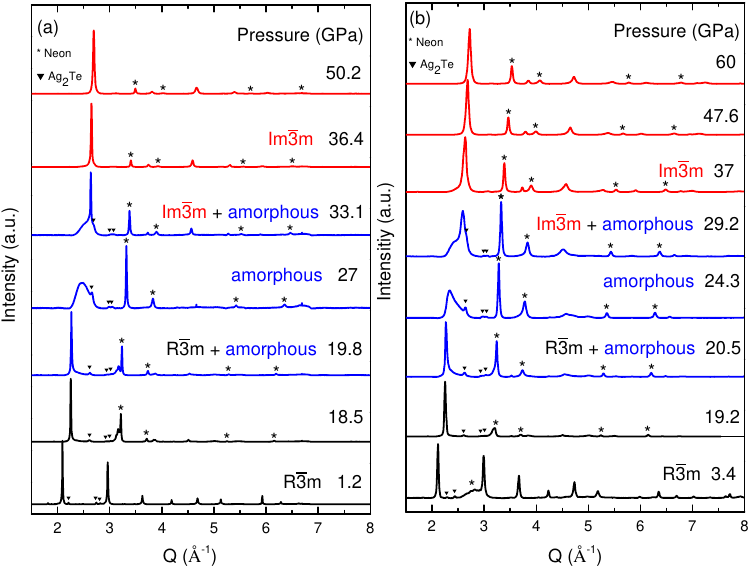}}
\centering
\caption{Selected XRD patterns, in scattering vector Q range, of AgSbTe$_2$ upon pressure increase, for two independent runs, (a) and (b). The x-ray wavelength is $\lambda=0.4142$\AA  {} and  $\lambda=0.2904$\AA {} in (a) and (b), respectively. The Bragg peaks originating from Ag$_2$Te (second minority phase) and Ne(PTM) are also denoted. A broad feature appearing in the XRD pattern at 3.4 GPa in panel (b), originates from liquid Ne (PTM). }
\end{figure}

This BCC-like phase remains stable up to, at least, 60 GPa.  A more detailed inspection of the XRD patterns just  before the amorphization at 18.5 GPa reveals the onset of a structural instability within the $R\overline{3}m$ phase. The main peak of $R\overline{3}m$ phase started to mismatch with the expected position,  while several other peaks split into two. Furthermore, additional peaks that are inconsistent with the $R\overline{3}m$ phase appeared, see Fig. S3. This effect was also reported by Kumar \textit{et al}. \cite{Kumar2005}, although the exact crystal structure was not provided. The vicinity of this structural instability with the onset of amorphization, also precluded us to accurately resolve the crystallographic details for this structure.

Preferred orientation and anisotropic broadening  effects precluded the use of Rietveld refinements in this study. From the relevant Le Bail refinements, see Figs. S4 and S5, of the XRD patterns of this study, the lattice parameters and the cell volume for the  AgSbTe$_2$ crystalline  phases were   determined and  are shown in Fig. 2, together with the corresponding calculated EoS.  Third-order Birch-Murnaghan EoSs \cite{Birch1947} were fitted to the  experimental data, and the determined bulk moduli are listed in Table S1   together with the corresponding lattice parameters at selected pressure. For the $R\overline{3}m$ phase, the  Bulk moduli determined in this study are practically  identical with ones determined by Kumar \textit{et al.}, for the B1 phase, that was assumed as the ambient conditions phase based on initial reports (see introduction).  The existence of the intermediate amorphous phase precludes a direct determination of the volume difference between the $R\overline{3}m$ and $Im\overline{3}m$ phases at the same pressure.  Nevertheless, it is plausible to assume, based on the extrapolation of the corresponding EOSs, that the overall volume difference is negligible. This is further supported by the V$_{p.f.u.}$ of the BCC phase upon pressure release, which, due to considerable hysteresis, remains stable at pressures lower than the onset pressure of the BCC phase and is practically identical to that of the $R\overline{3}m$ phase.

\begin{figure}[ht]
{\includegraphics[width=0.8\linewidth]{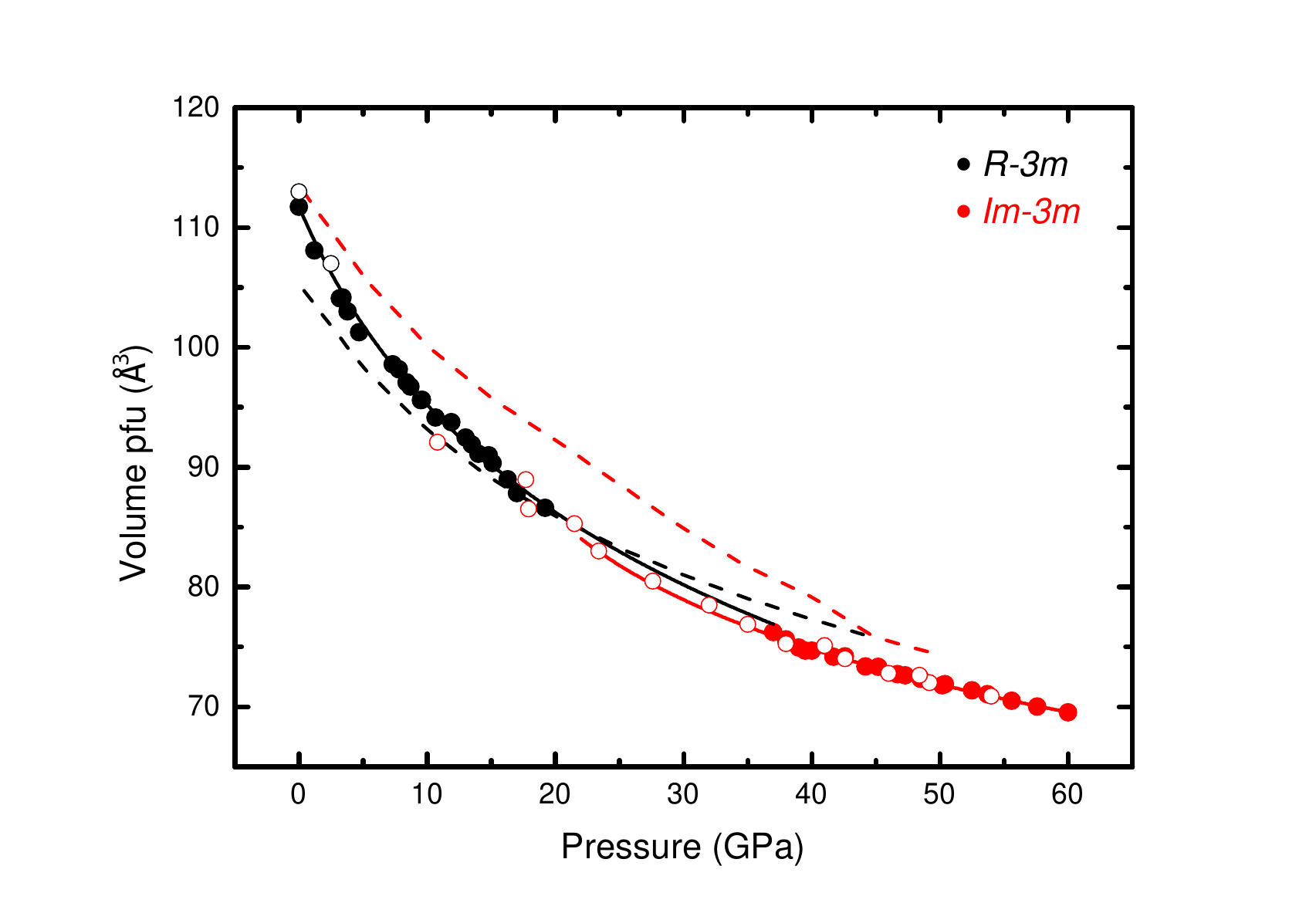}}
\centering
\caption{Pressure dependence of the  volume per formula unit  for AgSbTe$_2$. The solid lines are the EoS fitted to the experimental data (solid symbols) and the dashed lines are the EoS calculated by DFT.  The $R\overline{3}m$ and $Im\overline{3}m$ phases are indicated by black and red solid (open) symbols for increasing (decreasing) pressure, respectively. }
\end{figure}

Selected XRD patterns of AgSbTe$_2$ during decompression are shown in Figs. 3(a),(b) and (c), for three independent runs. Figs. 3(a) and (b) correspond to the same two experimental runs depicted in Fig. 1(a) and (b), respectively. In the first experimental run, when pressure was released gradually, as shown in Fig. 3(a), the $Im\overline{3}m$ phase remains stable  down to, at least, 10.8 GPa. Further decompression resulted in a transition to an  amorphous state. Few additional Bragg peaks originate from the $P2_1/c$ phase of Ag$_2$Te \cite{Sun2023}, that become apparent because of the amorphous phase of AgSbTe$_2$, see Fig. S1(b).  However, in the second experimental run, when the pressure was decreased directly from 17.5 to 2.5 GPa as shown in Fig. 3(b), the initial $R\overline{3}m$ phase was partially recovered, bypassing the amorphous phase,  with a, as normally expected, slight  broadening of the Bragg peaks  at ambient condition. Due to the relatively large scanning angle, a signal originating from the cubic boron nitride (c-BN) used as the seat of diamond was also observed in the diffraction pattern. To further investigate this kinetic effect more comprehensively, a third experimental run was conducted, in which the pressure was abruptly decreased directly from 35 GPa to ambient conditions. Remarkably, the $Im\overline{3}m$ phase was fully transformed  to the initial $R\overline{3}m$ phase with a high degree of crystallinity, see Fig. 3(c).

\begin{figure}[ht]
{\includegraphics[width=\linewidth]{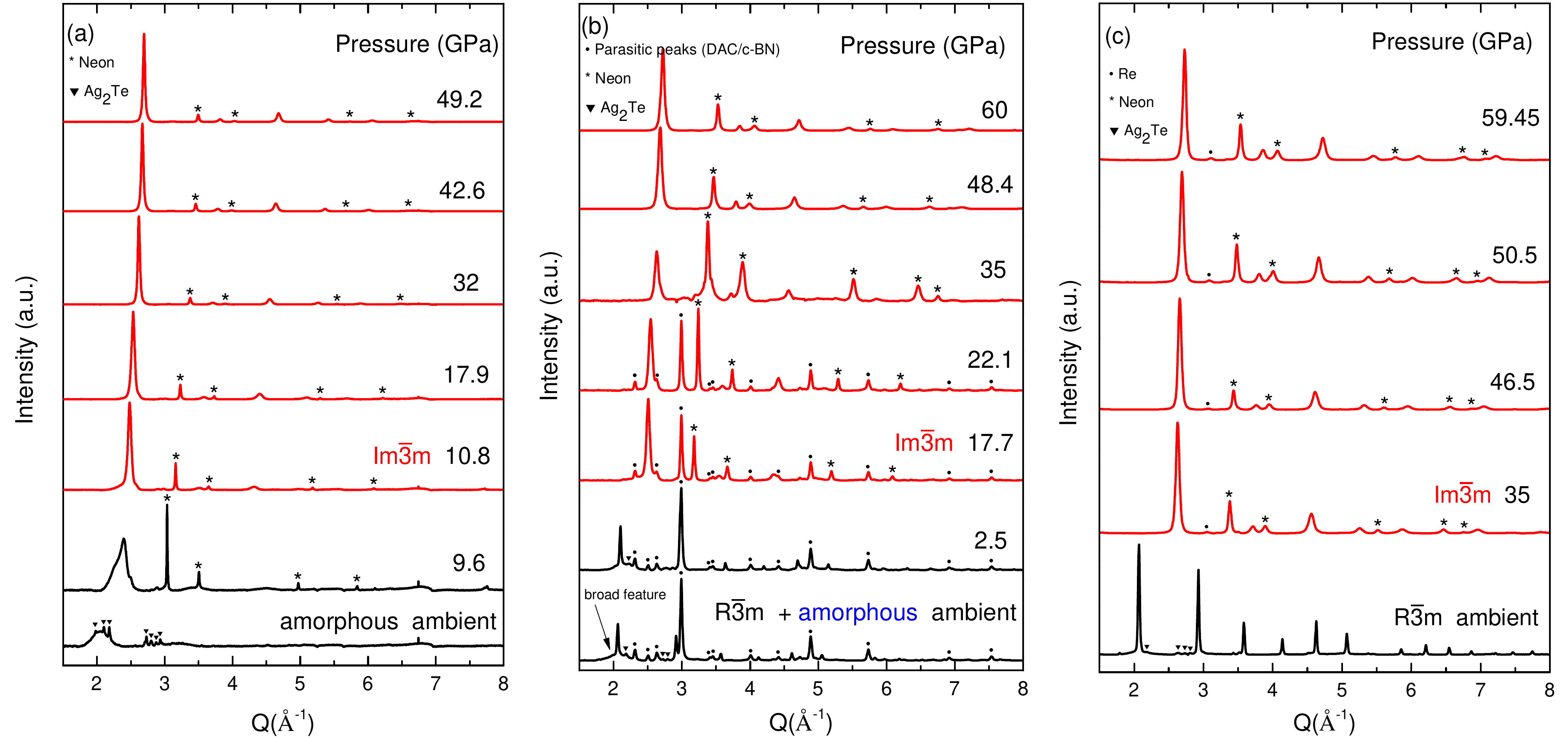}}
\centering
\caption{Selected XRD patterns, in scattering vector Q range, of AgSbTe$_2$ upon pressure decrease, for three independent runs, (a), (b) and (c). The x-ray wavelength is $\lambda=0.4142$\AA {} for (a) and  $\lambda=0.2908$\AA {} for  (b) and (c). The Bragg peaks originating from Ag$_2$Te (second minority phase), Ne(PTM), Re (gasket) and from DAC materials  are also denoted.}
\end{figure}

\subsection{Density functional theory calculations}

In the DFT calculations, the $R\overline{3}m$ phase was used as the ground state structure. To model the experimentally observed high-pressure phase, and to overcome the theoretical difficulty in simulating a disordered structure \cite{Zhu2011},  we constructed a partially disordered $Pm\overline{3}m$ 2 x 2 x 2 superstructure with Ag and Sb atoms occupying the 1a site (each with 0.5 average occupancy) and Te atoms fully occupying the 1b site. In Fig. 2, the calculated EoSs for both the  $R\overline{3}m$ and the $Pm\overline{3}m$ (effectively $Im\overline{3}m$) phases are presented and compared with the experimentally determined. Both experiment and theoretical EoSs show that the volume difference during the  phase transition is negligible, which points towards a second-order phase transition. 

The relative enthalpy difference for the two phases as a function of pressure is shown in Fig. 4. As expected, the $R\overline{3}m$ phase has lower enthalpy from ambient pressure up to $\approx$ 26 GPa. Above this pressure, $Im\overline{3}m$ becomes energetically favorable. However, the enthalpy difference does not increase monotonically with pressure. We postpone the discussion  about the implications of the pressure dependence of the relative enthalpies  to the observed PIA and the subsequent  recrystallization  for the discussion section.

\begin{figure}[ht]
{\includegraphics[width=0.8\linewidth]{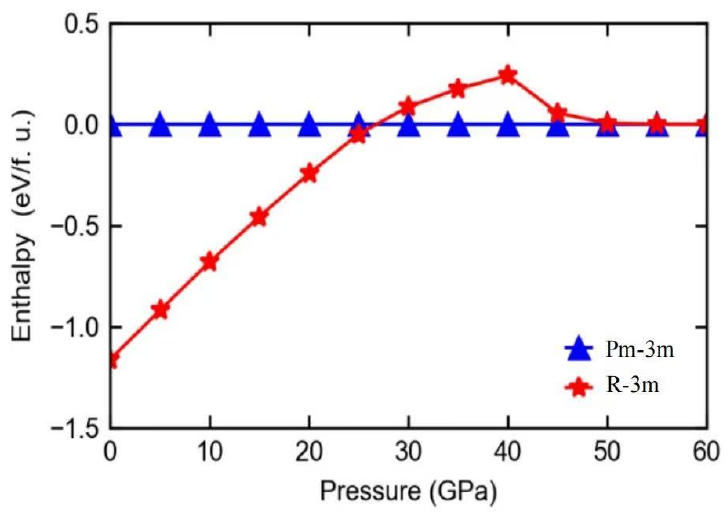}}
\centering
\caption{Relative enthalpy of the $R\overline{3}m$ and $Pm\overline{3}m$ ($Im\overline{3}m$)  phases as a function of pressure.}
\end{figure}

To study the mechanism of the pressure-induced  transition from the $R\overline{3}m$ phase to the $Im\overline{3}m$ phase, we further investigated the evolution of the local order of the $R\overline{3}m$ phase as a function of pressure. Fig. 5 shows the evolution of Ag–Te and Sb–Te bond lengths for the first and the second-nearest neighbours with pressure. As expected, the shortest Te-Ag and Te-Sb bonds (first-nearest neighbours) are more incompressible than the second-nearest neighbors  at pressure below 45 GPa. With increasing pressure, the bond differences rapidly decrease, especially above 20 GPa, eventually resulting in practically equal bond distances above 45 GPa. This strongly suggests that the  coordination number of the $R\overline{3}m$ phase increases at pressure above 45 GPa, from 6 to 8 in agreement with the experimentally observed BCC-like phase. Above the same pressure, the  c/a axial ratio, in the hexagonal representation,  and the rhombohedral angle of the $R\overline{3}m$ phase are close to the ideal values of   $\sqrt{6}$ and 60$^o$, respectively, as presented in Figs. S6 and S7.

\begin{figure}[ht]
{\includegraphics[width=0.8\linewidth]{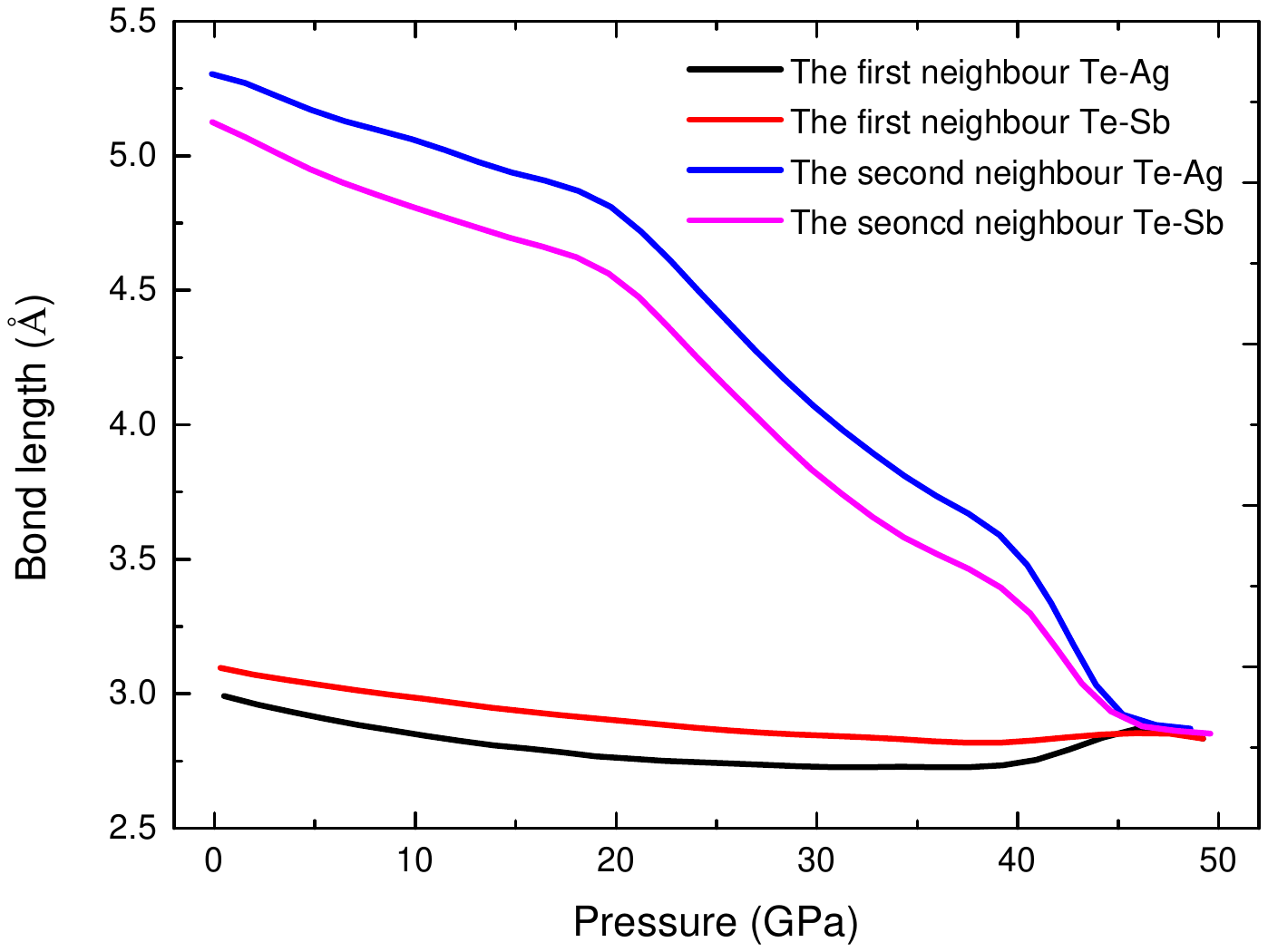}}
\centering
\caption{The first and second-nearest neighbors Te-Ag and Te-Sb bond lengths as a function of  pressure, calculated from DFT.}
\end{figure}

We further studied the symmetry change during the transition from  the $R\overline{3}m$ phase to $Im\overline{3}m$ phase. As shown in Fig. 6, Sb exhibits a -3m symmetry in the center of the rhombohedral cage of Te  for the $R\overline{3}m$ phase. As pressure increases, the shape of the Te cage changes from a rhombohedron to a cube, so the symmetry of Sb transforms from a -3m symmetry to a m-3m symmetry.

\begin{figure}[ht]
{\includegraphics[width=\linewidth]{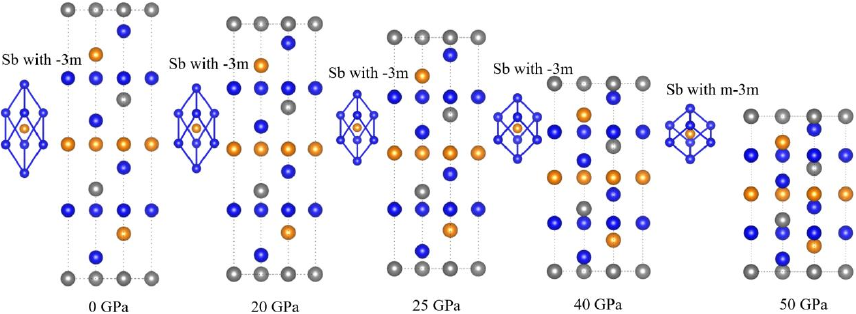}}
\centering
\caption{Crystal structure of the $R\overline{3}m$ phase of AgSbTe$_2$ at 0 GPa, 20 GPa, 25 GPa, 40 GPa and 50 GPa, upon pressure increase. The gray, orange and blue spheres denote Ag, Sb and Te atoms, respectively}
\end{figure}

\subsection{Molecular dynamics calculations}
MD simulations at different pressures starting from the   $R\overline{3}m$ upon pressure increase and from the  $Im\overline{3}m$ phase upon pressure decrease  are shown in Figs. 7(a) and (b), respectively. To clearly show the phase transitions from $R\overline{3}m$ and $Im\overline{3}m$, we highlighted the Sb-Te bonds and the Te polyhedra with central Sb as shown in Fig. 7. At ambient pressure, all the Sb atoms are located in the Te octahedra with a 6-fold coordination. As pressure increases, the  c axis quickly decreases and the coordination number of the Te atoms increases because of the emergence of Sb-Te bonds along the c axis. Notably, the increase of the coordination number of Te atoms progress  gradually with pressure, because only part of the Te atoms exhibit 8-fold coordination in simulation at 30 GPa as shown in Fig. 7(a). As a result, the phase transition from $R\overline{3}m$ to $Im\overline{3}m$ occurs gradually,  with a disordered displacement of the Te atoms progressing over a wide pressure region, in agreement with experimental observations. Likewise, the final structure, starting from the   $Im\overline{3}m$ phase at 50 GPa  and upon pressure decrease, when the pressure is fully released  is an amorphous phase because the Te polyhedral units were broken and did not revert to 6-fold coordination of the Te octahedra in $R\overline{3}m$. 

\begin{figure}[ht]
{\includegraphics[width=0.8\linewidth]{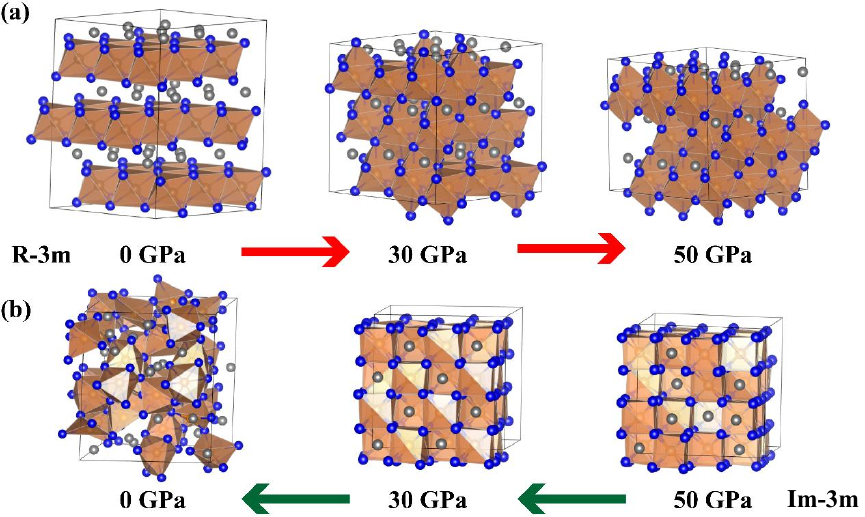}}
\centering
\caption{(a) The final structures derived from MD simulations upon pressure increase with the ordered $R\overline{3}m$ phase of AgSbTe$_2$ as the starting phase at 0, 30 and 50 GPa . (b) The final structures from MD simulations upon pressure release with the $Im\overline{3}m$ phase of AgSbTe$_2$ as the starting phase at 50, 30 and 0 GPa. The gray, orange and blue balls denote Ag, Sb and Te atoms, respectively. The Sb-Te bonds and Te polyhedra with central Sb are shown in the structures.}
\end{figure}

Further  insight was acquired from the pressure evolution of the corresponding radial distribution function (RDF) as a function of pressure. As pressure increases to 30 GPa, the characteristic RDF peaks of $R\overline{3}m$  phase are becoming wider  as shown in Fig. 8(a). On the other hand, the corresponding peaks of the total RDF  of $Im\overline{3}m$, shown in Fig. 8(b), become narrow above 30 GPa,  in agreement with the observed phase transition. The displacement of Te atoms was identified as the potential  origin of the phase transition by DFT calculations. To clarify this,   we also investigated the partial  PDF of the Te-Te bonds, that are shown in Fig. S8.  In the case of the $R\overline{3}m$ phase upon pressure increase (Fig. S8(a)), the corresponding   first neighbor Te-Te bond PDF peak is  splitting into 2 peaks and 3 peaks at 10 and 30 GPa, respectively. Likewise, the characteristic peaks of the corresponding Te-Te PDF  for the $Im\overline{3}m$ phase upon pressure release from 50 GPa (Fig. S8(b)) become broader, eventually resulting to a single broad feature upon full pressure release. On the other hand, the corresponding PDFs of the Sb-Te (Fig. S9), the Ag-Te (Fig. S10) and Ag-Sb (Fig. S11) bonds, do not show any displacement disorder, $i.e.$ splitting and exhibition of amorphization. All these observations signify the central role of Te atoms displacements on both the observed PIA upon applying  pressure to the $R\overline{3}m$ phase and also to the observed PIA upon pressure release starting from the $Im\overline{3}m$, independently of the presence  of vacancies. 

\begin{figure}[ht]
{\includegraphics[width=0.8\linewidth]{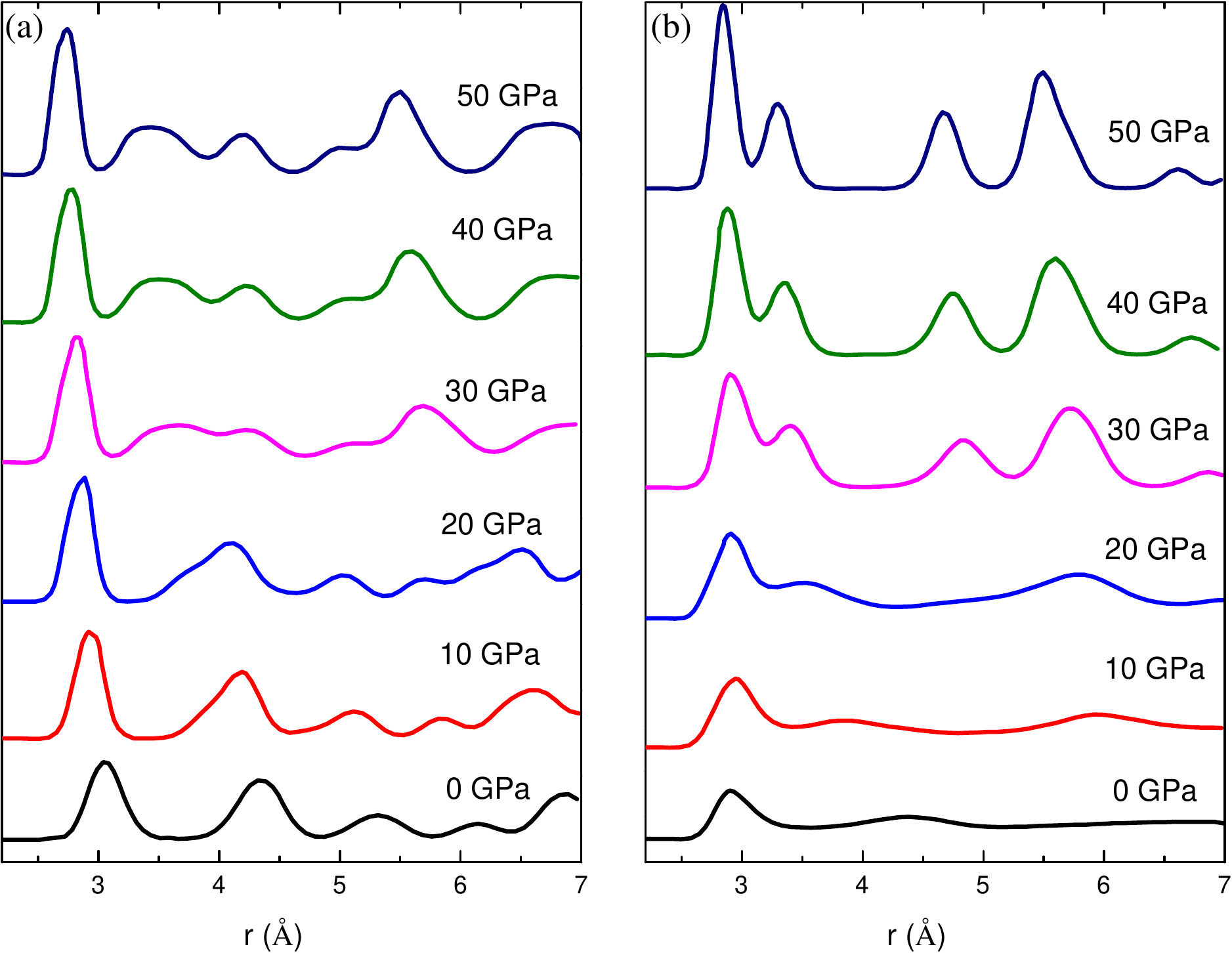}}
\centering
\caption{Total RDF of the (a) $R\overline{3}m$ upon pressure increase and (b) $Im\overline{3}m$ upon pressure decrease.}  
\end{figure}

\section{Discussion }
\subsection*{Pressure induced amorphization}
A PIA was experimentally observed at $\approx$ 20 GPa, with the amorphous phase remaining stable up to, at least, $\approx$ 37 GPa. The slight discrepancy of the critical pressures for the corresponding  phase transitions between this and previous studies can be explained based on the difference on the level of hydrostaticity.  Indeed, in this study Ne that is used as PTM remains fairly quasi-hydrostatic up to at least 50 GPa \cite{klotz2009}. On the other hand,  both silicon oil or 4:1 methanol: ethanol mixture used previously as PTMs \cite{Kumar2005,Ko2014} become substantially non-hydrostatic above 10 GPa \cite{klotz2009}. A non-hydrostatic PTM is known to have strong effect on the phase transitions critical pressures, see $e.g.$ Refs. \cite{zhang2023,shen2017,yang2021}. 

As mentioned in the results section, the enthalpy difference between the $R\overline{3}m$ and $Im\overline{3}m$ phases remains relatively small and with a varying trend ($i.e.$ first increasing and then decreasing) over a wide pressure range, pointing towards, effectively, an enthalpy degeneracy between the two structures.   This observation is in agreement with the experimentally observed amorphization, in the context that the enthalpies of both  phases  are so close that the system becomes amorphous (see detailed discussion in Ref. \cite{Ferlat2012}), before the final transformation to the high-pressure $Im\overline{3}m$ phase. It should be noted that at around 40 GPa, the enthalpy difference between the $R\overline{3}m$ and $Im\overline{3}m$ phases reaches its maximum within the 26–60 GPa pressure range. This indicates that at this pressure there should be a strongest thermodynamic driving force for the transformation from the $R\overline{3}m$   to the $Im\overline{3}m$ phase. This is in agreement with our experimental observation of the pressure induced recrystallization at approximately 37 GPa. However, based solely on enthalpy calculations, one might expect the system to continue to be amorphous, or will amorphize again as the enthalpy difference decreases from 40 GPa and approaches zero at 60 GPa. This interpretation, however, only reflects thermodynamic considerations and not barrier effects. Our decompression experiments reveal that the $Im\overline{3}m$ structure can revert to the initial $R\overline{3}m$ structure only when the decompression rate is sufficiently fast (thermal effect, see below discussion). With slow decompression, the $Im\overline{3}m$  can be even stable down to 10.8 GPa, highlighting a substantial pressure hysteresis, indicative of   a significant kinetic barrier for the reverse transformation, and thus stabilizing the $Im\overline{3}m$ phase. Therefore, once the phase transformation is completed (~37 GPa experimentally, corresponding to the maximum enthalpy difference near 40 GPa), the $Im\overline{3}m$  phase becomes kinetically stabilized.

Certain conclusions can be made about the driving cause of the observed PIA based on the combination of the DFT and MD calculations of this study. First, DFT calculations indicate an enthalpy degeneracy between the $R\overline{3}m$ and the $Im\overline{3}m$ phases above 25 GPa, accompanied by a substantial disordered displacement of Te atoms above 20 GPa as concluded by MD calculations. Second, the rates of the decrease, as a function of pressure, of the second nearest neighbor Te-Ag and Te-Sb bond lengths were increased at $\approx$ 20 GPa where is the starting point of the PIA.  Therefore, the large displacement of the Te atoms is concluded to be the origin of the amorphization during compression, irrespective of the presence of defects. This is further supported by the substantial broadening of the corresponding total and partial (Te) RDFs upon pressure increase. We note that the slight difference between experiment and theory on the onset and completion pressures of the amorphous state can be attributed to the fact that  calculations are performed after structural relaxation.

\subsection{High pressure phase characterization}
As mentioned in the introduction, a $Pm\overline{3}m$ (B2) structure,  where both Ag and Sb occupy the 1a site with 0.5 occupancy  and Te occupies the 1b site, was reported by Kumar \textit{et al}. \cite{Kumar2005} as the highest-pressure phase. However, no clear experimental evidences about the exact crystal structure were provided. Mainly, if this BCC-like phase is partially (B2, CsCl-type) or fully (A2, W-type) disordered.  The XRD patterns of this study  are in favor of a fully disordered A2 phase. Indeed,  we have not observed the characteristic Bragg peak of the $Pm\overline{3}m$ phase that should appear at $2\theta$  $\approx$ 5.1$^{\circ}$ as shown in the inset of Fig. S12. Thus, the high-pressure phase has the $Im\overline{3}m$ (A2) symmetry, $i.e.$   is fully disordered with all elements occupying the 2a WP with relevant partial occupancies (0.25, 0.25 and 0.5 for Ag, Sb, and Te, respectively). 

To examine whether elevated  temperature could result in an atomic ordering of the A2 phase, as in the case  of Bi$_2$Te$_3$ \cite{loa2016}, additional high-temperature experiments were performed.  For this reason, laser heating (LH) experiments were performed at 60 GPa up to 1700 K, the highest temperature before the decomposition into Ag$_2$Te and Sb$_2$Te$_3$ has been observed. The diffraction patterns obtained during high-temperature (HT) conditions and after quenching to ambient temperature exhibit no significant differences beyond peak-broadening and peak-shifts attributable to thermal expansion effects in the case of the XRD patterns under HT conditions.  The comparison between the experimental diffraction pattern at 60 GPa after LH with the simulated two possible BCC-like phases is shown in Fig. S12. The results experimentally support the fully disordered  $Im\overline{3}m$ phase, in contrast to the previously reported $Pm\overline{3}m$ phase \cite{Kumar2005}.

\subsection{Kinetic effect upon decompression}
The  decompression rate imposes a strong  kinetic  effect  on the structure of the quenched phase. Specifically, Figs. 3(a–c) demonstrate that as the decompression rate increases, the final phase after full pressure release evolves from (a) amorphous to (b) partially crystallized and eventually to (c) fully crystallized. Earlier studies proposed two distinct pathways: a thermally activated crystal-to-crystal transition under slow decompression, and a mechanically driven amorphization under rapid decompression \cite{Lin2020}. In contrast, our results show the opposite behavior: rapid decompression induces reversibility towards  the initial crystalline structure, whereas slow decompression favors amorphization. Detailed MD simulations, that are correlated  with experimentally imposed  slow decompression, in Fig. 7(b) revealed that the ambient conditions $R\overline{3}m$ phase   cannot be recovered after decompression from 50 GPa in agreement with experimental observations. Based on above observations, we argue that such counterintuitive  behavior should be related to the intrinsic thermal properties of the material. 

 During slow decompression, the system possesses sufficient time to explore kinetic pathways with lower energy barriers. This gradual relaxation process maintains the  amorphous intermediate state which is consistent with the amorphous region observed between the $R\overline{3}m$ and $Im\overline{3}m$ phases during compression. In contrast, experimentally imposed rapid decompression promotes reversibility due to the poor thermal conductivity (in contrast to the case of Si \cite{Lin2020}) of AgSbTe$_2$. In this scenario, the fast decompression ($i.e.$  non-equilibrium process) increases the temperature of AgSbTe$_2$,  effectively promoting reversibility towards the crystalline state. This is  in agreement with the relatively low temperature (75$^o$C) reported for the crystallization of amorphous AgSbTe$_2$,  \cite{Detemple2003}.  We note that Kumar \textit{et al}. \cite{Kumar2005} reported a crystallization of the  amorphous phase, recovered upon slow decompression, upon heating to 100$^o$C, further supporting our argument. This unusual kinetic effect displays behavior contrary to previously reported rate-dependent decompression mechanisms.

\section{Conclusion}
The high-pressure structural evolution of AgSbTe$_2$ has been investigated by a combination of XRD measurements and DFT and MD theoretical calculations. The results clearly document a pressure-induced phase transition from the ambient conditions  rhombohedral  $R\overline{3}m$ phase to a fully disordered cubic $Im\overline{3}m$ phase through an intermediate amorphous state, that is mainly associated to the pressure-driven substantial displacement of the Te atoms. 
Interestingly, a strong kinetic effect was observed in our study upon pressure release. Slow decompression result to an experimentally observed  amorphous phase, further supported by MD calculations. On the other hand, fast decompression results to a fully reversible transformation to the ambient conditions  $R\overline{3}m$ phase.

\section{Methods}
\subsection*{AgSbTe$_2$ speciment}

The same material as in Ref. \cite{sun2025} was used in this study.  More information about materials synthesis can be found in Ref. \cite{sun2025}. As described in Ref. \cite{sun2025}, a small amount ($<$5\%) of $\beta$-Ag$_2$Te \cite{Sun2023} is present as second, separated, phase. The minuscule quantity (see Fig. S1(a)) of this second phase, does not affect the high-pressure behavior of AgSbTe$_2$. 

\subsection*{High pressure X-ray diffraction studies}

Mini BX-80 and symmetric diamond anvil cells (DACs) with 300 $\mu{m}$ diamond culets were used for the high-pressure experiments. Rhenium gaskets were pre-indented to a thickness of 30 - 40 $\mu{m}$  followed by laser drilling of 100 $\mu{m}$ diameter holes at their centers. Ruby and gold were used as pressure markers, with ruby fluorescence  used to determine the initial pressure after gas loading \cite{piermarini1975}, while an equation of state (EoS) of gold provided in-situ pressure monitoring during high-pressure experiments at synchrotron radiation sources \cite{Anderson1989}. Neon  was used as the pressure transmission medium (PTM), maintaining quasi-hydrostatic condition up to the highest pressure of this study \cite{klotz2009}.

A Dectris Pilatus3 S 1M Hybrid Photon Counting detector was used at the Advanced Light Source (ALS), Lawrence Berkeley National Laboratory, Beamline 12.2.2. The  X-ray probing beam was focused to a spot size of 10 $\times$ 10 $\mu{m}$ using Kirkpatrick-Baez mirrors. More details on the XRD experimental setups are given in Kunz \textit{et al.} \cite{Kunz2005}. At SPring-8, beamline BL10XU, a Flat Panel X-ray Detector (Varex Imaging, XRD1611 CP3)  was used and the X-ray probing beam  was focused to about   10 $\times$ 10 $\mu{m}$  using compound refractive lens. More details on the SPring-8 XRD experimental setups are given in  Hirao  \textit{et al}. \cite{Hirao2020}.  At Beamline P02.2 at DESY, the X-ray probing beam was focused to a spot size of 2 $\times$ 2 $\mu{m}$ at the sample using Kirkpatrick-Baez mirrors and a PerkinElmer XRD 1621 flat-panel detector. Double-sided CW laser heating was performed using a focused IR (1064 nm) laser beam  to $\approx$ 10 $\mu{m}$ or 30 $\mu{m}$ in diameter (FWHM) spot at DESY and ALS,respectively,  which have online laser heating capabilities. Temperature was measured spectroradiometrically simultaneously with XRD measurements with a typical uncertainty of 150 K. 

Integration of powder diffraction patterns to yield scattering intensity versus 2$\theta$ diagrams and initial analysis were performed using the DIOPTAS program \cite{Prescher2015}. Calculated XRD patterns were produced using the POWDER CELL program \cite{Kraus1996} for the corresponding crystal structures according to the EoS determined experimentally in this study and assuming continuous Debye rings of uniform intensity. The fitted EoS to the experimental data was done by EoSFit7 GUI software \cite{GonzalezPlatas2016}. Indexing of XRD patterns has been performed using the DICVOL program \cite{Boultif2004} as implemented in the FullProf Suite. Le Bail refinements were performed using the GSAS - II software \cite{toby2013}. 

\subsection{Theoretical calculations}

\subsubsection{Density functional theory}
$Ab-initio$ DFT calculations were carried using the Vienna  simulation package (VASP 6) \cite{kresse1996}. The exchange-correlation functional is specified as the generalized gradient approximation \cite{perdew1996generalized1} parameterized by Perdew, Burke, and Ernzerhof \cite{perdew1996generalized2}. For high pressure EoS and phase transition, the valence electrons of the 4\textit{p}$^{6}$4\textit{d}$^{10}$5\textit{s}$^1$, 4\textit{d}$^{10}$5\textit{s}$^{2}$5\textit{p}$^{3}$ and 4\textit{s}$^{2}$5\textit{p}$^{6}$4\textit{d}$^{10}$5\textit{s}$^{2}$5\textit{p}$^{4}$ orbitals were considered for the elements Ag, Sb and Te, respectively. The plane wave cutoﬀ energy was set to 600 eV, and the Monkhorst-Pack k-spacing value was selected to be 0.03 2$\pi$/\AA. The structural optimizations were carried out from 0 GPa to 60 GPa with a pressure step of 5 GPa. At each pressure, the crystal structures are fully relaxed, where ionic positions, cell volume, and cell shape of the structures are allowed to change in optimizations.

\subsubsection{Molecular dynamics}
$Ab-initio$ MD simulations were carried  at 0, 10, 20, 30, 40 and 50 GPa using on-the-fly machine learning force fields \cite{jinnouchi2019}, where the 3 $\times$ 3 $\times$ 1  and 2 $\times$ 2 $\times$ 2 supercells were used for ordered $R\overline{3}m$ phase and $Fm\overline{3}m$ phase, respectively. An ordered $Fm\overline{3}m$ superstructure was used to model the  disordered solid solution phase $Im\overline{3}m$.
Valence electrons of the 4\textit{d}$^{10}$5\textit{s}$^{1}$, 5\textit{s}$^{2}$5\textit{p}$^{3}$ and 5\textit{s}$^{2}$5\textit{p}$^{4}$ orbitals were considered for the elements Ag, Sb and Te, respectively. The plane wave cutoﬀ energy was set to be 400 eV, and the k-mesh was  gamma-centered. Langevin NPT thermostat was used to examine temperature at room temperature (300 K) with a  simulation time of 30 ps (time step of 1 fs) at each pressure, allowing thermal equilibrium at each pressure.

\subsection{Data Availability}
The data that support the findings of this study are available from the corresponding author upon reasonable request.

\bibliography{AgSbTe2pressure}

\subsection{Acknowledgments}
The work performed at GTIIT was supported by funding from the Guangdong Technion Israel Institute of Technology and the Guangdong Provincial Key Laboratory of Materials and Technologies for Energy Conversion, MATEC (Grant No. 2022B1212010007, Guangdong Department of Science and Technology). We thank the financial support from the Technion-GTIIT seed grant program. Z.Z., W.L. and R.A. acknowledge the support from the Swedish Research Council (VR) (Grant No. 2020-04410). The computations were enabled by resources provided by the National Academic Infrastructure for Supercomputing in Sweden (NAISS), partially funded by the Swedish Research Council through grant agreement 2020-04410. Y.A. would like to acknowledge generous support from the Pazy Research Foundation, Grant No. 398. Beamline 12.2.2 at the Advanced Light Source is a DOE Office of Science User Facility under contract no. DE-AC02-05CH11231. Beamline 12.2.2 is partially supported by SEES: SEES is supported by the National Science Foundation Division of Earth Sciences (EAR) SEES: Synchrotron Earth and Environmental Science (EAR –2223273). Part of the synchrotron radiation experiments were performed at BL10XU  of SPring-8 with the approval of the Japan Synchrotron Radiation Research Institute (JASRI) (Proposal Nos. 2024A1096 and 2024B1182). We acknowledge DESY (Hamburg, Germany), a member of the Helmholtz Association HGF, for the provision of experimental facilities. Parts of this research were carried out at PETRA III beamline P02.2. We thank the Shanghai Synchrotron Radiation Facility beamlines  BL17UM (31124.02.SSRF.BL17UM) and BL15U1 (31124.02. SSRF. BL15U1) for the assistance on the high-pressure XRD measurements. We thank Sergei Grazhdannikov and  Yaron Amouyal for providing the AgSbTe$_2$ specimen. 

B.S. and  E.S. conceived the experimental part of the project and designed the experiments.  Z.Z., W.Luo. and R.A. conceived the theoretical part of the project and designed the calculations. W.Lu., S. F., H. Z., C.W. and A.S.A. performed experimental measurements and analysis. M. K., H. K. and B. W. provided methodology resources and experimental support. S. G. and Y.A.   synthesized the specimens.  B.S., E.S., Z.Z. and R. A. wrote the
manuscript with comments from other authors. All authors reviewed the manuscript.

\subsection{Competing interests}
The authors have no conflicts to disclose.

\end{document}